\documentclass[12pt]{article}
\begin{document}
\title{Quantum Evolution as a Square Root of the Master Equation}
\author{J.M.J. van Leeuwen}
\date{}
\maketitle
\begin{center}
{\it Instituut-Lorentz, Universiteit Leiden,\\
Niels Bohrweg 2, 2333 CA Leiden, The Netherlands.}
\end{center}

\begin{abstract}
  The analogy between the quantum evolution and that of the master
  equation is explored. By stressing
  the stochastic nature of quantum evolution a number of conceptual
  difficulties in the interpretation of quantum mechanics are
  avoided. 
\end{abstract}

\section{Introduction}

Quantum mechanics is hundred years old. As with many revolutions the
precise birthdate is not clear, but the definite breakthrough from
classical mechanics to quantum mechanics was the article by Heisenberg
in 1925 \cite{heisenberg}. In this paper it became clear that the
quantum theory was
not a set of rules added to classical mechanics but a completely
new mechanics, which forced us to think differently about the micro world.
Quantities like position and impulse of particles
changed from having definite values to matrices or operators in
Hilbert space.
After the invention of classical mechanics by Newton,
quantum mechanics is considered to be the greatest discovery
in physics with the 
largest impact on our view on the nature of matter. With the
subsequent probability interpretation of Born \cite{born}, the
deterministic classical dynamics was replaced by a quantum
stochastic process. 

In particular the stochastic character of quantum dynamics
met with resistance. Einstein objected to this aspect with
`` Der lieber Herr Gott w\"urffelt nicht'' and asked for an underlying
deterministic theory. However such a deterministic foundation is
not compatible with the structure of the {\it Heisenberg-Born} formulation
of quantum mechanics as von Neumann \cite{neumann} showed. Even today
one struggles with the proper appreciation of the quantum
process as weird notions like the ``many worlds'' and ``wave function
of the universe'' demonstrate.

An informal poll \cite{tegmark} taken in July 1999 at a conference on
quantum computation at the Isaac Newton Institute in Cambridge, England,
suggests that the prevailing viewpoint is shifting.
Out of 90 physicists polled, only eight declared that their view
involved explicit wave-function collapse.
Thirty chose many worlds or consistent histories (with no collapse).
As stated in \cite{tegmark} the picture is not clear: 50 of the
researchers chose none of the above or were undecided. 
Rampant linguistic confusion
may contribute to that large number. It is not uncommon for
two physicists, who say that they subscribe to the Copenhagen
interpretation, for example, to find themselves disagreeing
about what that means to them. This is also my personal experience in
talking to fellow physicists about the interpretation of quantum mechanics.

A link between quantum mechanics and classical stochastic processes
is not new. Already in 1933 Furth \cite{furth} investigated
the similarities between the classical stochastic processes and
quantum mechanics.
However this paper did not attract much attention and it was
recently translated and reissued by Peliti and Muratore-Ginanneschi
\cite{peliti}. 
Furth constructed uncertainty relations for the
stochastic diffusive process in analogy with the uncertainty relations
in quantum mechanics. A more bold step was made by Nelson \cite {nelson} by
constructing a classical stochastic process that obeyed the
Schr\"odinger equation.
  
In this chapter I am less
ambitious and explore the similarities and differences between
a classical stochastic process and the evolution of a quantum process.
I will show that many of the misunderstandings
about the quantum process can
be avoided if one exploits the analogy between the quantum process 
and a classical stochastic process, in particular Brownian motion.
From this analogy one obtains a better insight in the quantum
process than the often-used notion of particle-wave duality. On the
other hand it is also interesting to elaborate on the fundamental
differences between quantum and classical stochastic processes.
The essential point in this analysis is the distinction between the
stochastic process and the evolution of the probability distribution.
In order to make this point let us start with the analogy between
Brownian motion and the Schr\"odinger equation for a free particle.

\section{Brownian motion and the Schr\"odinger Equation}
Brownian motion is governed by the diffusion equation
\begin{equation} \label{a1}
\frac{\partial P({\bf r})}{\partial t} = D  \Delta P({\bf r},t),
\end{equation}
where $P({\bf r}, t)$ is the probability of finding the particle at
position $\bf r$ at time $t$. $D$ is the diffusion coefficient of the
Brownian particle. The equation needs an initial distribution
$P({\bf r}, 0)$ for the complete solution $P({\bf r}, t)$. The
typical solution starts with a delta function $\delta(\bf r)$ in
the origin. The corresponding $P(\bf r, t)$ is then given by\footnote{
  We follow the convention that the vector ${\bf r}$ is given by
  a boldface character and its length $|{\bf r}|$ by a lower case
  symbol $r$,}
\begin{equation} \label{a2}
  P({\bf r}, t) = \frac{1}{(4 \pi D t)^{3/2}} \exp[- r^2/(4 D t)]
\end{equation}
Note that the equation conserves probability
\begin{equation} \label{a3}
  \int d {\bf r} P({\bf r}, t) = 1.
\end{equation}
The evolution equation is deterministic, since the initial value of $P$
determines the $P({\bf r}, t)$ at later times.

If the particle is observed to be at $\bf r'$ at time $t'$,
the new starting point for the probability distribution is
\begin{equation} \label{a4}
  P({\bf r}, t') = \delta ({\bf r - r'}).
\end{equation}
In quantum mechanical language one would say that
through observation, the probability distribution ``collapses''
from a wide distribution to a delta function. Obviously the collapse
is not something physical since the probability
distribution $P({\bf r},t)$ is not a property of the Brownian
particle; it only gives the probabilities of finding the position
upon observation. The deterministic evolution of the probability
distribution between two measurements of the position does not
imply that the Brownian particle has moved deterministically
from the first found position to the later found position.

By fourier decomposition, one notes that all the eigenvalues of the
diffusion operator are negative except one.
All the wave components with finite 
eigenvalues decay and the one with eigenvalue zero is stationary.
To phrase the physics of the Brownian particle in a formulation akin
to the quantum mechanical: the Brownian particle is in a state
which can be found by measurement of its position.
The evolution equation gives the transition probabilities from
state to state. One can only measure displacements of the Brownian
particle; it has no velocity.

Now compare this with a free particle obeying the Schr\"odinger equation
\begin{equation} \label{a5}
\frac{ \partial \Psi ({\bf r},t)}{\partial t} =  
\frac{i \hbar}{2m} \Delta \Psi({\bf r},t),
\end{equation}
The similarity between the classical equation (\ref{a1}) and
the quantum mechanical equation (\ref{a5}) is
striking, but the association of the probability with $\Psi$ is
different. Not $\Psi$ but $|\Psi|^2$ is conserved
\begin{equation} \label{a6}
  \int d {\bf r} |\Psi ({\bf r} ,t)|^2 =
  \int d {\bf r} |\Psi ({\bf r} ,0)|^2.
\end{equation}
In spite of the formal correspondence of the Eqs. (\ref{a1}) and
(\ref{a5}), one cannot just copy the solution (\ref{a2}) to
$\Psi({\bf r},t)$. The reason is that the resulting $\Psi$ is not
normalizable in the sense of Eq.~(\ref{a6}). It is interesting however
to start with a near $\delta({\bf r})$ of the form
\begin{equation} \label{a7}
  \Psi({\bf r}, 0) = \frac{1}{(\pi \alpha)^{3/4}} \exp(-r^2/2 \alpha)
\end{equation} 
Taking the square of this function yields a $|\Psi|^2$ which is
normalized to 1. For small $\alpha$ the probability distribution is
confined to a narrow region around $ r = 0$. Via fourier
transforms one can solve Eq.~(\ref{a5}) leading to
\begin{equation} \label{a8}
  |\Psi({\bf r}, t)|^2 = \left(\frac{\gamma}{\pi}\right)^{3/2}
  \exp(-\gamma r^2),
\end{equation}
with $\gamma$ given by
\begin{equation} \label{a9}
  \gamma = \frac{\alpha}{\alpha^2+\hbar^2 t^2/m^2}.
\end{equation}
The larger $t$ is, the  smaller $\gamma$ will be, and the more the
probability distribution is spread out over space.

This example shows that the deterministic evolutions of the probability
distributions of the two processes are similar,
but involve different domains: the classical Brownian
motion refers to real positive probability numbers, while the quantum
Schr\"odinger equation operates in the plane of complex probability
amplitudes. The common feature is that an observation interrupts the
probability evolution and provides a new starting point for the
probability distribution. It is important to note that the
deterministic evolution of the probability is not the same as the
stochastic evolution of the particle. For the Brownian particle this
is obvious, but for the quantum particle the (deterministic)
Schr\"odinger equation is often considered as the evolution of the
particle while it gives only the evolution of the probability
amplitude. As in the Brownian case the quantum particle evolves in
a stochastic process. So one may not conclude from the deterministic
evolution of the probability amplitude between two interruptions
by the measurements that the quantum particle has moved in a
deterministic way from one event to another.

It may seem that the similarity between Brownian motion and
the evolution of a quantum free particle is coincidental. In the
following section we show that a much more general similarity exist
between the master equation and the Schr\"odinger equation.

\section{The Master and Schr\"odinger Equations}

The master equation provides the stochastic evolution of variable $X$.
It is defined by the transition rates $W(X|X')$ of state 
$X'$ to $X$. The probability $P(X,t)$ of state $X$ then obeys the equation
\begin{equation} \label{b1}
\frac{\partial P(X,t)} {\partial t} =\sum_{X'}
\left[ W(X|X') P(X',t) - W(X'|X) P(X,t) \right]
=\sum_{X'} H_{X,X'} P(X',t).
\end{equation}
The first term gives the gain to state $X$ and the second term the loss
from state $X$.
The equation is deterministic, i.e.~the probability at any later time $t$ 
is determined by the probability at an earlier time (e.g.~$t=0$). 
To be somewhat more precise, $P(X,t)$ gives the conditional
probability of finding the system in state $X$ at time $t$ given
the probability distribution $P(X,0)$ at time $t=0$.
In this discrete formulation the time evolution results
from a matrix operation on the distribution. The matrix $H_{X,X'}$ 
is called {\it stochastic} since the sum over the columns vanishes, 
as (\ref{b1}) shows. 
This property guarantees the conservation of probability 
\begin{equation} \label{b2}
\sum_X P(X,t)=\sum_X P(X,0),
\end{equation}
implying that one of the eigenvalues of $H_{X,X'}$ is zero; the others must
have a negative real part, otherwise the probability distribution
would grow without limit.

The general Schr\"odinger equation has the form
\begin{equation} \label{b3}
i \hbar \frac{ \partial \Psi (X,t)}{\partial t} =  
{\cal H} \, \Psi(X,t),
\end{equation} 
where $\Psi$ is the wave function (state) of the system. 
This is also a deterministic equation, since the initial value of the 
wave function $\Psi(X,0)$ determines the behavior at later times. 
The hamiltonian operator $\cal H$ is hermitian. For classical systems
there are strict rules for deriving the Hamiltonian from the
Hamilton function by replacement of position and impulse by operators. 
Equation (\ref{b3})  yields the conservation law
\begin{equation} \label{b4}
\int dX |\Psi (X,t)|^2 = \int dX |\Psi (X,0)|^2,
\end{equation} 
which is interpreted as the conservation of probability.

More generally one sees $\Psi$ as a state $|\Psi \rangle$  and
$\Psi(X)$ as the inner product of the eigenstate of the operator $X$
and $\Psi$
\begin{equation} \label{b5}
  \Psi(X) = \langle X| \Psi \rangle.
\end{equation} 
The probability
$P(X)$ of finding the value $X$ in state $\Psi$ is given by
\begin{equation} \label{b6}
  P(X) = |\langle X | \Psi \rangle |^2.
\end{equation}
After measuring $X$ the system is in the state $|\Psi\rangle=|X\rangle$,
where $X$ is the result of the measurement.
Thus, the quantum process has an evolution equation and a rule for the
probabilities. This defines a stochastic process. The important
difference from the master equation is that the probability is the
absolute square of that value of $P(X,t)$ that obeys the evolution equation.
Another difference is that $P(X,t)$ not only leads to the probability of
finding $X$,  but it also contains, by transformation of the basis, the
probability of other observables as e.g. the momentum.

The quantum problem is much richer than the
stochastic process of the master equation. In the Appendix the
correspondence is worked out between a Brownian particle in a
harmonic force field and the ground state of a quantum particle in
a harmonic well. Only the ground state
distribution can be matched with a solution of the master equation.
The excited states of the quantum particle in the harmonic potential
have no counterpart in the solutions of the master equation. The
picture of a quantum particle that evolves according to a stochastic
process like a Brownian particle explains the zero point motion in
the lowest energy state of the quantum particle. While the probability
distribution in the ground state is stationary the particle
continues to fluctuate.

\section{Similarities and Differences between the Processes}

As we pointed out the evolution of the probability distribution
refers in both cases to an underlying stochastic process. The
difference is that for the Brownian motion one can easily visualize
a deterministic motion that leads on a larger scale to a stochastic
motion. In the quantum case there is no room for a deterministic
motion on a smaller scale.

Another fundamental difference between the classical master
equation and the Schr\"odinger equation is that the latter
allows a complex  value of the probability
amplitude, while the probability in the master equation must remain
positive and real. So tunneling is not a real option in the master equation
Brownian motion, while complex probability distributions in the
Schr\"odinger equation make the phenomenon of tunneling possible. 
  
One may view the solution of the Schr\"odinger equation as a
square root of the solution of the master equation.
The stochastic character of both processes is the
same. The evolution of the process concerns the probability
distribution in the master equation, while it concerns a square root
of the probability in the Schr\"odinger equation.
Square roots are not innocent though.
To mention a few profound differences: the wave function is complex
and the time evolution is unitary and not decaying. So while a plane wave
decays in the master equation, it evolves via a phase factor in the
Schr\"odinger equation.

As the Schr\"odinger equation is linear one
can make superpositions of states, which gives an association with
waves. Therefore $\Psi$ is called the wave function, a somewhat
unfortunate name, as probability amplitude would have been better.

The analogy with stochastic processes is even stronger for many particles.
Probabilities for more than one particle are naturally defined in the 
product space of the particles. The probability in the case of
a pair of particles
is a function in a six-dimensional space. Similarly
the wave function of two particles 
is a function in a six-dimensional space. This shows all the more that the
solution of the Schr\"odinger equation is not a wave function but a
probability amplitude.

\section{The Two-Slit Experiment}

We take one of the experiments in quantum mechanics,
the two-slit experiment, in order to illustrate the differences
between the master equation and the Schr\"odinger equation. It became
famous as it was the object of the Bohr-Einstein discussion in the
Solvay meeting of 1927. 
An electron is fired at a barrier with two open slits.
After passing through the slits, the electron is detected at a 
screen. For the full experiment two detection devices observe which of
the two slits the electrons passes through. These detection devices
can be turned on and off.

According to the master equation the probability of the electron
arriving at the screen is the sum of the probabilities of  passing
through one or other of the slits. In both paths the probability
of arriving at a particular point of the screen is the probability
of passing through the relevant slit, multiplied by the probability
of passing from that slit to the point in question.
Since probabilities are positive, all points on the screen
have a positive probability of detecting the electron. If the
particle is detected at one of the slits the probability of reaching
the screen is the probability of that path. For the probability
distribution at the screen it is unimportant whether the particle is
observed while passing through one of the slits
 
However, for a quantum particle the screen displays an
interference pattern, since in quantum
mechanics one must add the wave function and not the absolute squares.
The interference pattern is reminiscent of coherent light passing through 
the two slits. No wonder that the analogy with waves has been invoked
and that the Schr\"odinger equation 
is seen as a {\it wave equation}. This is an unfortunate misnomer,
suggesting that
$\Psi$ is a wave of a field, as Schr\"odinger originally thought.
The analogy is misleading, since one can also detect the passing
of the electron through the slits.
The outcome is that the electron is always detected as passing through one
of the slits, never through both at the same time.
The detection of the electron at one of the slits destroys
the interference pattern. A wave, on the contrary, interferes with
itself, since it is half present at both slits.
The result for the electron is not strange if one realizes that 
quantum evolution is a 
stochastic process. So if the electron is found at one slit, one must 
continue with that new initial value for the wave function.

\section{Classical versus quantum}

A probabilistic description is meaningful if the probability
distribution has a non-zero variance.
If the variance shrinks the probability
distribution becomes a prediction.
Consider the trajectory of a charged elementary
particle in a cloud chamber. The condensation trail that the particle
leaves behind can be seen as a repeated measurement of the position
of the particle. Each droplet that is formed due to the passage of
the charge is a measurement of its position.
The trail of droplets is wide with respect to the size of the
particle. So the probability distribution of the position of the
particle shows a large variance. If we look on the other hand at the
trajectory of the moon, the quantum variance is negligble. One can
predict the trajectory of the moon many years ahead.

There is a gradual transition from quantum to classical behaviour.
It is not sharp as in a phase transition but it is a crossover.
Sufficiently far away from
the crossover region the distinction between micro and macro is
clear.

In the quantum mechanical description, there are parameters (involving
Planck's constant $h$) that tell
whether the system behaves classically or quantum mechanically. An example
is the de-Boer parameter $\Lambda$ for noble gases
\begin{equation} \label{d1}
  \Lambda = \frac{h}{\sigma \sqrt{m \epsilon}},
  \end{equation} 
indicating whether a noble gas is classical like
argon or quantum mechanical like helium. Here $\sigma$ is the size of
the hard core, $m$ is the mass of the particles and
$\epsilon$ is the depth of the interaction well.

There is no point in ascribing a quantum distribution to a
macroscopic object. Invoking the wave function of the universe makes
no sense and wrongly suggests that the time evolution of
this wave function has a relation to how the universe actually
develops.

\section{The Measurement Process}

Measuring the position of a quantum particle is in essence not different
from measuring the position of a Brownian particle. In both cases, one
wants to know where the particle is. However measuring a quantum
object is more difficult than measuring a classical object. One reason is that
one must connect the micro world of the quantum system with the macro
world of the measuring device. So, it helps
if the measuring device is in a metastable state like a cloud chamber,
where the trail of a charged particle is recorded through the
irreversible formation of droplets. Another reason, which has
historically played a large role, is that the measuring device has an
influence on the quantum system. This was the fundamental issue in
the Bohr-Einstein debate. It is however possible to design
measurements with a smaller and smaller energy transfer, such that
the measurement device becomes a passive element for the quantum particle.
A third
reason is that quantum system plus measuring device must be well
isolated from external influences which is much more delicate for
quantum systems than classical systems. 

Measuring whether the electron in the two-slit experiment has passed a
slit requires a macroscopic change in the status of the measuring device.
If the device is fluctuating as a microscopic object, it is 
useless as a measuring device. Measuring a quantum observable
requires that the interaction of the observable with the device
is capable of changing the status of the device. The trail of an
elementary particle in the cloud chamber is a paradigm of the
interaction of a quantum particle with a macroscopic system. Each
droplet generated by the elementary particle is a measurement of
its position. The thickness of the trail is a measure for the
variance of the position of the particle. It is magnified by the
growing of the droplets. Such magnification is often helped
by a metastable
state of the device in which small effects can have large
consequences. It is of great importance that the probability
distribution of the electron does not depend on the details
of the measuring device but only on the state of the electron.

\section{Why is the Master Equation So Easy to Accept ?}

There is little discussion of the status of the (classical)
master equation. One of the reasons is that
the probabilities obey the standard rules of classical probability
theory. In particular the Kolmogorov theorem is obeyed, which states
that the probability of getting from the initial state $X_0$ at $t_0$
to the final state $X_2$ at $t_2$ equals the sum over the probabilities
of all paths via intermidiate states $X_1$ and time $t_1$. In the
two-slit experiment, the probability of the electron reaching a
particular on the screen
is the sum of the probabilities of passing through one or the other
to that point. This is in line with our intuitive
notion of probabilities.

Another reason is that the master equation is an approximate
description of the physical process.
One usually has an idea of the underlying mechanical
theory, and the master equation is invoked because the underlying
mechanics is too difficult. Another aspect showing the approximate
character of the equation is that the temporal evolution is
irreversible. All the eigenmodes of the master equation decay, except
the mode corresponding to the eigenvalue zero, which yields the
stationary state.
Although the macro world is obviously irreversible,
the micro equations have to be time-reversal-invariant. The master equation
applies to the mesoscopic behaviour: it is less detailed than the
micro description, but it still contains the fluctuations around the
macroscopic variables.

\section{Why is Quantum Mechanics So Difficult to Accept?}

The first clash between the two heroes of the old quantum theory,
Einstein and Bohr, concerned the two-slit experiment. Einstein
considered quantum mechanics incomplete since the electron sometimes
behaved like a particle (while passing through the slits) and sometimes
as waves (while interfering at the screen). So, he asked for an
underlying deterministic motion.
The answer from Bohr was
rather clumsy, using the particle-wave duality and the uncertainty
relations for the interaction between the electron and the measuring
device. Viewing the quantum process as a stochastic process with the
rules of forming the probabilities, the answer would have been clear and
unambiguous. Since Bohr was the major opponent of Einstein, the Born
rules came to be known as the Copenhagen interpretation. But Bohr was,
with his notion of particle-wave duality and later complimentarity,
not the ideal spokesman of the Born rules for the probability.

The stochastic character of quantum evolution is unavoidable, but hard
to swallow for a mind that has grown up with causality in the form:
every event has a cause. This does not hold in quantum
mechanics. If the electron passes in the two-slit experiment through
the upper slit, there is no reason why it has chosen the upper and
not the lower slit. The reasoning of Bohr, which led him to the notion
of complementarity is a dead-end street.
One simply has to accept that the outcome is
intrinsically unpredictable. Von Neumann \cite{neumann} had
already shown that the structure of quantum mechanics
does not permit hidden variables.

The dispute also triggered the question: ``What happens during the
measurement?'' ``Is the Schr\"odinger equation still valid?''.
All that happens is a transition of the state of the object
to the eigenstate of the observable, corresponding to the
found value. That is usually called a collapse of the wave function,
which is more misleading than explaining. A collapse suggests an action
of the measurement device on the wave function, while it concerns
only a new start for the probability amplitude. In Brownian
motion, it is clear that the Brownian particle is the reality and
the evolution equation only the description of the probability
distribution. The quantum case is the same. The fluctuating particle
is the reality and the Schr\"odinger equation the description
of the probability amplitude.

Still some give the Schr\"odinger equation the
status of being the exact mechanics of the wave function. 
Van Kampen \cite{kampen} aptly summarizes his
opinion  in his fourth theorem (with a slight modification):

{\bf Theorem IV.}  {\it Whoever endows the wave function with more meaning than 
is needed for computing transition probabilities is responsible for the 
consequences.}

The extreme interpretation of the wave function as a physical field
governing the evolution of any system is the idea
of the wave function of the universe.
One should keep in mind that quantum mechanics is only tested in
isolated systems and that only in isolated systems can meaningful
calculations be carried out. The marriage of the quantum evolution
with the probabilistic rules leads to the idea of many
worlds. Every time a choice is made, a new world is created. All other
possible choices create simultaneously parallel universes.
Apart from the fact that this idea is useless for calculations,
there is also no trace of
influences of the many worlds on each other.

The psychology of giving a higher status than a probability amplitude
to the solution of the Schr\"odinger equation is understandable.
Quantum mechanics replaced the classical deterministic mechanics.
The quantum equations are
mostly constructed from the classical equations of motion by replacing
variables such as position and momentum by operators in a well-defined way.
Later on operators appeared without a classical analogue such as spin
(which Dirac derived by studying the square root out of the
Klein-Gordon equation). 
As the hamiltonian is the basic ingredient in the evolution of the
wave function, all the experience with hamiltonians, 
such as conservation laws, can be exploited in the Schr\"odinger equation. 

Stochastic processes were, so far, known to be approximate.
One knew the underlying mechanistic motion of the Brownian 
particle. Moreover the master equation cannot be fundamental, since it
is not time-reversal invariant, while a fundamental
theory has to have this property. 
In fact, in quantum mechanics the situation is the opposite: the
Schr\"odinger equation  is time-reversal invariant and 
there is no underlying deterministic mechanics.

For a number of people, this is not a satisfactory answer. They hope to
get information on the measurement by studying the Schr\"odinger
equation for the combined system of object and measuring device.
This is an interesting mathematical exercise, but it will not give a clue
as to which of the possible answers is realized. For the probability
distribution of the measured value one does not need to know the
role  of the measuring device since the probability distribution
is given by the state of the observable.

While the discussion between Bohr and Einstein becomes irrelevant
once the stochastic nature of the quantum process is accepted, the
second attack on quantum mechanics by Einstein, Podolsky and Rosen
\cite{epr} is much more inspiring.
It involves a system of two particles which
are initially close together and interacting, and then fly apart. By
measuring on one of them, one can get information on the other
particle without the necessity that the first
signals to the second how it has to behave. The issue originally
referred to the question whether one could ascribe ``elements of
reality'' to a system without actually measuring them. Later
Einstein objected to  ``spooky interaction at a distance''.
The conclusion is that quantum mechanics, relativity
and locality do not go together.

In the later variant of the ``EPR paradox'' by Bohm \cite{bohm}, one
constructs, ``EPR pairs'', a singlet state of two particles, one with
spin up and one with spin down,
\begin{equation} \label{d1}
  | \psi \rangle = \frac{1}{2} ( | 1\uparrow, 2 \downarrow \rangle +
  | 1 \downarrow, 2 \uparrow \rangle ).
\end{equation} 
After the construction of the pairs, the constituting particles 1 and 2
are separated far apart. Then the spins are measured. There is a 50\%
probability that one finds particle 1 to have spin up or down
and the same holds for particle 2. Inspecting the records after the
measurements one finds that if 1 is up,
2 is down or vice versa. The separation of the particles is so
large and the times of measurement so close that no signal from 1
to 2 is possible to tell how 2 has to react to the measurement of 1.
This is Einstein's ``spooky interaction at a distance''. If we replace
``interaction'' with ``correlation'', it sounds already less spooky.
Particles 1 and 2 are indeed correlated or entangled in the
singlet state.

The discussion remained purely theoretical till
Bell \cite{bell} made the distinction between classical and
quantum probabilities more explicit. He constructed an inequality
holding for all possible classical
correlations and showed that in special cases quantum
correlations violate this classical inequality. This was confirmed
by the experiments of Aspect \cite{aspect}.
At present, one is searching to exploit
the features of entanglement. It cannot be used as a means of
transmiting information (if so John Bell hoped that they would call it
the ``Bell Telephone''). But entanglement can be used as a nonbreakable
encryption and as a check on eavesdropping. Also, for the operation
of the quantum computer entanglement is an essential feature.

The unitary evolution of the wave function
of a quantum system is denoted as {\it conservation of information}.
This becomes a big point in astrophysics: can quantum information
escape from a black hole? As soon as the word information is used,
the discussion becomes a bit vague. The sort of information  
carried by the wave function is nothing else 
than a way to calculate probabilities.

We take again the two-slit experiment for illustration. 
One would think that by measuring the passage at the slit, information is
gained on the electron. On the other hand, information on the electron
is destroyed since its wave function ``collapses''. I presume that the
proponents of the ``conservation of information'' will argue that the
combined system of electron and measuring device evolves unitarily,
so total information is conserved. But by unitary evolution, the combined 
system remains forever in a superposition of possible outcomes, 
while in reality only one happens. So, the sort of information carried
by the unitary wave function of the combination is at odds with reality.

Suppose that we performed a two-slit experiment with {\it classical} particles,
in which not only the arrival distribution at the screen is observed, but
also which slit the particle passes through. Then one has two options
to calculate the distribution at the screen: use the information
of detection at the slits or ignore it. Both lead to the
same result. Using it, one can calculate two sub-distributions related
each to a passage at one of the slits. 
The sub-distributions contain more information.
By Kolmogorov's theorem, their sum is the same as that calculated without 
knowledge of which slit was passed.

The situation in quantum mechanics is different. If the particles are
observed passing 
through the slits, no interference pattern appears on the screen. 
So, if the passing through the slits is measured but ignored (maybe
someone else measures them), one draws the wrong conclusion about
the probability distribution at the screen.
This is a clue to why eaves dropping shows up in 
transmission of quantum information and not in classical information. 
It prompts me to formulate the following theorem:   

{\bf Theorem IVa.} {\it Anyone ignoring information provided by
  measurements draws wrong conclusions.}

\section{Conclusion}

It is remarkable that even Einstein, one of the founding fathers of
Brownian motion theory, had such trouble accepting the stochastic nature
of quantum mechanics. Originally, he set up a stochastic interpretation,
but later got convinced that ``God does not play dice'',
to which Bohr replied: ``How do you know what God does?''

The quantum evolution equation for the probability amplitude
together with the Born rules for calculating the probabilities define
a stochastic process. Acknowledging this, quantum dynamics is crystal
clear. There is no room for an underlying deterministic mechanics.

Maybe the lingering quarrels with quantum mechanics have to do with
the traditional introduction for freshmen, which 
follows the historical part from the old quantum theory of Bohr
via particle-wave duality to the Schr\"odinger equation. The latter
has the advantage that it is easier to visualize as a wave equation
similar to the equation for electromagnetic waves.
The more fundamental approach of
Heisenberg is only introduced later.
This puts the students on the wrong track;
one should make students familiar with
stochastic processes before teaching them quantum mechanics.

This advice is against the present trend where the introduction to
quantum mechanics is already placed in the high-school curriculum.
Indeed quantum mechanics and relativity are the most spectacular theories
in physics and it is tempting to teach them in an early stage in the
curriculum. The price of this approach is that the students get the
impression that quantum mechanics is a mystery, while, in fact, the
rules of
quantum mechanics are clear. As consolation one then offers
them the quote from Feynman \cite{feynman}:
``nobody understands quantum
mechanics''.

{\bf Ackowledgements}

I am indebted to Michael Nauenberg, Henk Hilhorst,
Gerard Nienhuis and Dennis Dieks for their critical comments on the manuscript. Numerous discussions on the subject
with Bernard Nienhuis are gratefully acknowledged.

\section{A Particle in a Harmonic Potential}

A specific example of the master equation is the Brownian particle.
The state of the particle is the position ${\bf r}$. When the
transition probabilities
are confined to small steps, the master equation can be written in the
form of a Fokker-Planck equation. Introduce the step
\begin{equation} \label{cq}
{\bf s} = {\bf r} - {\bf r}'
\end{equation} 
and write the transition probability as
\begin{equation} \label{c2}
W({\bf r}|{\bf r'}) = W({\bf r'};{\bf s}).
\end{equation} 
Using this notation the master equation gets the form
\begin{equation} \label{c3}
\frac{\partial P({\bf r}, t)}{\partial t} = \int d {\bf s} \, 
W({\bf r}-{\bf s};{\bf s}) P({\bf r}-{\bf s},t) - P({\bf r},t) 
\int d {\bf s} \, W({\bf r}; - {\bf s}).
\end{equation} 
The assumption is that $W$ and the probability $P$ depend smoothly on the 
first argument ${\bf r}$. Then we may expand the first term in the master 
equation with respect to ${\bf r}$, while keeping the second argument
in its full glory
\begin{equation} \label{c4}
W({\bf r}- {\bf s}; {\bf s}) P({\bf r}- {\bf s},t) =
W({\bf r}; {\bf s}) P({\bf r},t) - [ {\bf s} \cdot \nabla] 
W({\bf r}; {\bf s}) P({\bf r},t) + \cdots
\end{equation}
Inserting this expansion in the master equation (\ref{c3}), the first term
in the expansion cancels the second term in the master equation. Thus the
master equation transforms into the Fokker-Planck equation
\begin{equation} \label{c5}
\frac{\partial P({\bf r},t)}{\partial t} = -\nabla \cdot {\bf a}_1 ({\bf r}) 
P({\bf r},t) + \frac{1}{2} \nabla \nabla \cdot {\bf a}_2 ({\bf r}) P({\bf r},t).
\end{equation} 
where the ${\bf a}_i ({\bf r})$ are the jump moments
\begin{equation} \label{c6}
{\bf a}_i ({\bf r}) = \int d  {\bf s} \, {\bf s}^i \, W({\bf r};{\bf s}). 
\end{equation}

The simplest jump moments are
\begin{equation} \label{c7}
  {\bf a_1} = - a_1 {\bf r}, \quad \quad \quad
  {\bf a}_2 ({\bf r}) = 2 D \, {\bf I}. 
\end{equation}
The first jump moment favors the jumps towards the origin, the more
the farther out the particle is. It represents the harmonic force
on the particle. The second jump moment takes as
lowest order the isotropic distribution. This gives the Fokker-Planck
equation the form
\begin{equation} \label{c8}
\frac{\partial P({\bf r},t)}{\partial t} = a_1\nabla \cdot  {\bf r}\, 
P({\bf r},t) + D \Delta P({\bf r},t).
\end{equation} 
The stationary state distribution has the form
\begin{equation} \label{c9}
  P({\bf r}) = \left(\frac{\kappa}{2 \pi} \right)^{3/2} \exp (-\kappa r^2/2).
\end{equation}
with $\kappa$ given by
\begin{equation} \label{c10}
  \kappa = a_1 /D.
\end{equation}

The Schr\"odinger equation for a particle in a harmonic potential
$ b \, r^2/2$ reads
\begin{equation} \label{c11}
  \i \hbar \frac{\partial \Psi({\bf r},t)}{\partial t} = - \frac{\hbar^2}{2m}
  \Delta \Psi({\bf r},t) + \frac{1}{2} b \, r^2 \Psi({\bf r},t).
\end{equation}
The stationary states are the energy eigenstates obeying the equation
\begin{equation} \label{c12}
  E \Psi ({\bf r}) = - \frac{\hbar^2}{2m} \Delta \Psi ({\bf r}) +
  \frac{1}{2} b \, r^2 \Psi({\bf r}).
\end{equation} 
The ground state is again a Gaussian
\begin{equation} \label{c13}
  \Psi_0 ({\bf r}) = \left(\frac{\lambda}{2 \pi} \right)^{3/4}
  \exp (-\lambda r^2/4),
\end{equation}
with the relations
\begin{equation} \label{c14}
  E_0 = \frac{3 \hbar^2 \lambda}{4 m} \quad \quad \quad
  b = \frac{\hbar^2 \lambda^2}{4 m}.
\end{equation}
The second relation gives the value of $\lambda$ in terms of the
harmonic force parameter $b$.
If we match the ground state distribution with the probability
distribution of the stationary state of the Fokker-Planck equation
\begin{equation} \label{c15}
  \Psi_0 ({\bf r}) ^2 = P({\bf r}),
\end{equation} 
we have to set $\kappa=\lambda$ and have to choose $a_1$ as
\begin{equation} \label{c16}
  a_1 = D \kappa = D \lambda =
  D \left( \frac{4 m b}{\hbar^2} \right)^{1/2}.
\end{equation}

\end{document}